\documentstyle[prl,aps]{revtex}
\begin{document}
\draft
\author{V.I. Valtchinov $^{(a,b)}$, G. Kalman $^{(a)}$ and K.B. Blagoev $^{(a)}$}
\title{Phases in Strongly Coupled Electronic Bilayer Liquids}
\address{$^{(a)}$ Department of Physics, Boston College, Chestnut Hill, MA 02167, USA\\
$^{(b)}$ Department of Radiology, Brigham and Women's Hospital, \\
Harvard Medical School, Boston, MA 02115, USA\\
}
\date{\today}
\maketitle
\begin{abstract}
The strongly correlated liquid state of a bilayer of charged particles has
been studied via the HNC calculation of the two-body functions. We report
the first time emergence of a series of structural phases, identified 
through the behavior of the two-body functions.
\end{abstract}


Electronic bilayer systems, consisting of two quasi-two-dimensional layers
of electron (or hole) liquids, separated by a distance $d$ comparable to the
interparticle separation $a$ (the Wigner-Seitz radius) within the layers,
have revealed an unexpected richness of features and have been the object of
many recent investigations. While most of the interest has originated from
the quantum Hall effect, and thus the related studies have been directed at
bilayers in strong magnetic fields, more recently it has been recognized
that unmagnetized bilayers also constitute a physical system of remarkable
complexity. Electronic bilayer systems in semiconductors can now be quite
routinely fabricated by modern nano-technology in a fairly wide range of
densities (i.e. $r_s$ values) and interlayer separations. Interestingly,
virtually similar systems have come into existence also under quite
different circumstances in ionic traps, where the charged particles are
classical ions and the layers develop as a result of the system seeking its
minimum energy configuration\cite{D91,TB88,GBW88}.

At high enough $r_s$ ($=a/a_B$, $a_B=\frac{\varepsilon _b\hbar ^2}{m^{*}e^2}$
is the effective Bohr radius) or $\Gamma $ ($=\frac{e^2}{akT}$ ) values
(estimated to be between $\Gamma =98$ and $\Gamma =222$\cite{GP95} ) the
bilayer is expected to crystallize into a Wigner lattice. While there is no
direct experimental evidence on the formation of such an electronic solid in
semiconductor bilayers without magnetic field (since the required $r_s$
values do not seem to be attainable by present-day experimental technique),
the formation of layered crystalline structures has been observed in
molecular dynamics computer simulations\cite{TB88} and in ionic traps\cite
{GBW88}. At the same time, a series of rather thorough theoretical
investigations \cite{D91,GP95,F94,EK95} have been carried out
predicting the formation of classical Wigner lattices in electronic or ionic
bilayers. These studies have revealed the existence of\ distinct structural
phases, whose characteristics depend on the layer separation, i.e. on the $%
d/a$ ratio: (see inset in Fig. 2): the staggered rectangular (ii); 
the staggered rhombic (iv), and the staggered triangular (v). Structure
(ii) comprises the simple triangular (i) and the
staggered square (iii), as special cases. A similar quantum
mechanical study for the magnetized bilayer\cite{NH95} has led to very
similar predictions.

The understanding of the experimentally more important (and theoretically
perhaps more challenging) {\it liquid phase}, on the other hand, is quite
incomplete. Its analysis hinges upon the availability of the {\it intralayer}
and {\it interlayer} two-body functions, ($g_{11}(r)$ and $g_{12}(r)$,
respectively). While attempts have been made to determine the $g_{ij}$-s
through various approximations, none of the results obtained can be
considered reliable especially with regard to correctly reflecting the
developing short range order: Zheng
and McDonald \cite{ZM94} and Neilson and collaborators\cite{NSSL93}, 
have used the
STLS (Singwi-Tosi-Land-Sjolander) approach, whose limitations are
well-known; Kalman, Ren and Golden\cite{KRG93} have used a quasi-iterative 
method, which breaks down for small layer separations. 

The purpose of this Letter is to provide the first
reliable calculation of the pair correlation functions $g_{ij}(r)$ and
related quantities such as the structure functions $S_{ij}(k)$ and the
dielectric matrix $\varepsilon _{ij}(k)$ for a bilayer system. The results
show that different short range ordered structures develop already in the
liquid phase, at relatively modest coupling values, emulating the different
structural phases in the Wigner lattices. The gradual increase of the layer
separation is accompanied by dramatic changes in $g_{ij}(r)$, $S_{ij}(k)$
and $\varepsilon _{ij}(k)$ that can be partially interpreted in terms of the
structural changes referred to above, partially in terms of the new type of a
{\it substitutional} orded-disorder transition. The bilayer seems 
to be a unique liquid system exhibiting structural transformations 
of the phase pursuant to the
change of a system parameter: such transformations are commonly known only
in the {\it solid} phase.

Our method of approach is based on the classical HNC (Hyper-Netted Chain)
approximation. This method has proven to be extremely reliable and accurate
for Coulomb systems, both in three dimensions\cite{RYDR83,T78} and two
dimensions\cite{T78,GCC79,L78}. We have adopted the HNC method to the
bilayer situation, by mapping the (single component) bilayer onto a
two-component $2D$ system, with an interaction potential matrix\cite{L78,VKB}
\begin{equation}
\varphi _{11}=\varphi _{22}=\frac{2\pi e^2}r;\varphi _{12}=\varphi _{21}=%
\frac{2\pi e^2}{\sqrt{r^2+d^2}};
\end{equation}
and using the two-component equivalent of the HNC method. The two layers are
assumed to have equal densities $n$. The resulting system of equations for 
$g_{ij}(r) = h_{ij}(r)+1$, $c_{ij}(r)$, and their Fourier transforms 
$h_{ij}(k)$ and $c_{ij}(k)$ 
\begin{equation}
\FL
h_{ij}(r)=e^{h_{ij}(r)-c_{ij}(r)-\phi _{ij}(r)}-1;
h_{12}(k) =\frac{c_{12}(k)}{[1-nc_{11}(k)]^2-[nc_{12}(k)]^2};  
h_{11}(k) =\frac{c_{11}(k)+nc_{12}(k)h_{12}(k)}{1-nc_{11}(k)};
\end{equation}
has been solved following Lado's original work on $2D$ systems\cite{L78}.
The underlying model is a classical one, where the average kinetic energy is
represented by the inverse temperature parameter $\beta $ and the intralayer
coupling by the parameter $\Gamma =\frac{e^2}{akT}$ . This description is
appropriate for bilayers arising in ionic traps if the confining potential
is steep enough so that the degree of freedom perpendicular to the bilayer
planes is suppressed; it is expected to be also quite adequate for the
analysis of the spin-independent structural phases of the zero-temperature
electron liquid, in the domain of high enough coupling ($a>a_B$) where the
electrons are quasi-localized and the effect of intralayer exchange becomes
small. Interlayer tunneling effects also disappear under the usually less
stringent $d>a_B$ condition. Under these circumstances, invoking the equivalence 
$\Gamma \simeq 2r_s$ is a reasonable prescription for the interpretation of
the results derived from the present model of semiconductor bilayers.

Our result for the intralayer $g_{11}(r)$ and interlayer $g_{12}(r)$ two-body 
functions are summarized in Figs.1 which show their variation as
functions of the coupling $\Gamma $ and of the interlayer separation $d/a$.
In order to quantitatively analyze the information contained in the results
we have used as principal indicators:

(i) $g_{12}(0)$, the measure of the probability of finding a particle in
layer $2$ above another particle in layer $1$: when the interlayer
correlations become significant, $g_{12}(r)$ approaches zero;

(ii) $R_I$, $R_{II}$, $R_{III}$ and $Q_I$, $Q_{II}$, $Q_{III}$, the
positions of the first three maxima of $g_{11}(r)$ and $g_{12}(r)$,
respectively: these maxima identify the correlation-shells around a given
particle, both in its own layer and in the layer adjacent to it;

(iii) the intralayer coordination numbers $\rho _I$, $\rho _{II}$, $%
\rho_{III}$ and the interlayer coordination numbers $\sigma _I$, $\sigma
_{II}$, $\sigma _{III}$, associated with each shell; the coordination
numbers are defined by $\rho _i=2n\pi \int_{M_i^{^{\prime }}}^{M_i^{^{\prime
\prime }}}rg_{11}(r)dr$, $\sigma _i=2n\pi \int_{N_i^{^{\prime
}}}^{N_i^{^{\prime \prime }}}rg_{12}(r)dr$ where $M_i^{^{\prime }}$ ($%
N_i^{^{\prime }}$) and $M_i^{^{\prime \prime }}$($N_i^{^{\prime \prime }}$)
are the minima preceding and following $R_i(Q_i)$.

Short range order, signaled by the onset of oscillating behavior for $%
g_{11}(r)$ appears around $\Gamma _s$ $=3$, more or less the same value as
for the isolated $2D$ layer. The value of $\Gamma _s$ doesn't seem to be
significantly affected by the layer separation $d$. Once the short range
order has developed, the most dramatic novel feature of the two-body
functions is the series of quite abrupt shifts in the positions and changes
in the amplitudes of the first few peaks both of $g_{11}(r)$ and of $%
g_{12}(r)$. This can be well observed in Figs. 1 and 2. These features
unambiguously point at the formation and phase-transformation-like changes
of short range structures in the bilayer electron liquid, which can be
paralleled with the results of recent studies of the bilayer crystalline
phase\cite{D91,GP95,F94,EK95}.

The physical features of the
system are revealed through the two-body functions,
the structure functions and the dielectric matrix of the system. In order to
bring out the characteristic features in the
details of the structure of two-body functions we
have used the rather high $\Gamma =80$ value for the latter. For the
structure functions and for the dielectric matrix we have used the more
realistic $\Gamma =20$ and $\Gamma =10$ values.

1. {\it Two-body functions} $g_{11}(r)$, $g_{12}(r)$.

1.1 Interlayer correlations, as inferred from $g_{11}(0)$ seem to completely
disappear for about $d/a=3$, even for very high $\Gamma $ values 
(see inset Fig. 2). Thus for layer separation higher than this 
value the two layers behave as virtually independent $2D$ systems.

1.2 The positions, heights, and widths, and the resulting coordination
numbers of the correlation-shells shift and change drastically as the
interlayer separation is increased (see Fig. 2). These changes can be
brought into correspondence with the expected formation of the different
lattice structures in the solid phase as discussed above. Since each lattice
structure carries a precise set of nearest neighbor positions $\{R_\alpha \}$%
, $\{Q_\alpha \}$ and coordination number $\{\rho _\alpha \}$, $\{\sigma
_\alpha \}$(with $\alpha $ $=1,2,3...$ representing the first, second,
third, etc. nearest neighbors in the lattice), these can be identified with
the previously defined $\{R_\alpha \}$, $\{Q_\alpha \}$, $\{\rho _\alpha \}$%
, and $\{\sigma _\alpha \}$ (values with $\alpha =I,II,III$, representing
the first, second and third shell positions). (The identification is not
one-to-one: a given shell, in general, with the possible exception of the
first shell, contains more than one set of nearest neighbors). 
Thus the formation of a series of ''liquid phases'' emulating the 
underlying lattice structural phases can be observed.

1.3 At $d=0$, when the two layers are collapsed into a single $2D$ system
with double density $2n$, their two two-body functions $g_{11}(r)$ and $%
g_{12}(r)$ are identical ,with $R_{II}/R_I=Q_{II}/Q_I\simeq 1.9$ ratio,
slightly higher than the characteristic $\sqrt{3}$ of a triangular lattice.
The $R_I=Q_I$ $=1.25$ value itself is somewhat below the lattice constant $%
R_1=1.34$, similarly to what has been obtained in the Monte Carlo and HNC%
\cite{T78,GCC79,L78} calculations for an isolated $2D$ electron liquid. The
double density triangular structure can also be regarded as a staggered
rectangular lattice with a side ratio $a_2/a_1=\sqrt{3}$. The identity of
the two two-body functions $g_{11}$ and $g_{12}$ and the accompanying
equality of the coordination numbers, $\rho _i=\sigma _i$, however, clearly
indicates that the two species are in a complete {\it substitutional disorder%
}, occupying the lattice sites at random.

1.4 With $d$ increasing away from the zero value, the substitutional {\it %
disorder} is rapidly replaced by a substitutional {\it order} appropriate
for the staggered rectangular structure: by $d/a=0.5$, $\rho _I$ and $\sigma
_I$ assume their expected $\rho _1=2$ and $\sigma _1=4$ values. At the same
time, the rectangular unit cell deforms toward a square-like shape: this is
well shown by the reduction of $R_{II}/R_I$ to the $\simeq 1.5$ value. The
substitutional disorder that prevails in the domain $0<d/a<0.5$ would seem
to require that vestiges of the peaks of $g_{11}(r)$ show up in $g_{12}(r)$
and {\it vice versa}. This, however, is apparently {\it not }the case: $R_I$
and $Q_I$, $R_{II}$ and $Q_{II}$, etc. are quite distinct. Nevertheless, the
separations between $R_i$ and $Q_i$ remain small enough and the peaks wide
enough to accommodate the substitutionally ill-positioned particles. In
fact, the $Q_{III}$ shell appears to be thriving on the ill-positioned
population: by the establishment of the substitutional order the $II$-shell 
at $Q_{II}$ $=3.65$ vanishes and is replaced by the more distant
position at $Q_{III}$ $=4.99$.

1.5 The transformation to an underlying staggered square lattice structure
becomes complete at $d/a=0.7$. The indicators of this transformation are the
jump of $R_{II}$ from $2.30$ to $3.89$ and $R_{II}$ from $3.87$ to $5.63$,
and the accompanying jump of the coordination numbers $\rho _I$ from $\simeq
2$ (same as $\rho _1$ for the rectangular lattice) to $\simeq 7$ (close to $%
\rho _1+\rho _2=8$ ($\rho _1=4,\rho _2=4$) for the square lattice). All
these changes reflect the higher symmetry of the square lattice.

1.6 In the domain $d/a>0.7$ the staggered square lattice gradually
transforms into a staggered {\it rhombic} lattice with $90>\varphi $ $%
>60^{\circ }$: this allows the energetically favorable increase of the
lattice constant, at the cost of the {\it decrease} of the energetically
less important second and third neighbors. This trend is very clearly
observable in the behavior of $R_I$, $R_{II}$, and $R_{III}$ and also
detectable in $Q_I$, and $Q_{III}$. There is no significant change in the
coordination numbers in this domain, except a slight reduction in $\rho _I$, 
$\rho _{II}$, $\sigma _I$, $\sigma _{II}$ reflecting the reduced distance
between the corresponding shells.

1.7 Around $d/a=1.6$, $\varphi $ reaches the $60^{\circ }$ value and in each
layer the underlying structure becomes a standard $2D$ triangular
(hexagonal) lattice, with lattice constant appropriate for density $n$. This
can be inferred from $R_I$ reaching its maximum and $R_{II}$, $R_{III}$ and $%
Q_I$, $Q_{II}$, $Q_{III}$ reaching their minimum values, the former very
close to $\sqrt{2}$ times the $R_I$($d=0$), $R_{II}$($d=0$) and $R_{III}$($%
d=0$) values respectively.

1.8 The triangular lattices emerging as the end product of the rhombic
structure are staggered in such a way that the vertices of layer $2$ lie
over the midpoints of the sides of the triangles in layer $1$ (and vice
versa). This is not the minimum energy configuration for well-separated
layers: this latter is attained when the vertices of the $2$-triangles lie
over the centers of the $1$-triangles. Transition to this final
configuration, consisting of a rigid translation of lattice $2$ with respect
to lattice $1$, takes place in the domain $1.5<d/a\simeq 2.0$ and is shown
by a slight, but perceptible increase in $Q_I$ and $Q_{III}$.

2. {\it Structure functions} $S_{11}(k)$, $S_{12}(k)$ Fig. 3). 
The expected small $k$
behavior of the structure function is governed by a generalized
compressibility sum rule\cite{KGunp}: at $k=0$ the $S_{11}(0)=-S_{12}(0)$
condition is required; both of the values of $S_{11}(0)$ and of 
$\frac{\partial S_{11}(k)}{\partial k}(k=0)$ are governed by the 
quantity $L_{11}-L_{12}$, the difference
between the inverse compressibility and the inverse trans-compressibility: $%
L_{ij}=\frac{\partial P_i}{\partial A_j}K_o$, where $K_o$, $P$ and $A$ are
respectively the compressibility of the non-interacting gas, the pressure
and the surface area. At $d=0$ the purely correlational $L_{12}$ compensates
the correlational contribution to the $L_{11}$ (resulting in $%
S_{11}(0)=-S_{12}(0)=1/2$). This is also the consequence of the perfect
screening sum rule which requires that $n\int d^2r\{h_{11}(r)+h_{12}(r)\}=-1$%
. For high $d$ values $L_{11}$ becomes dominant, approaching the value
appropriate for the isolated 2D layer, which, for $\Gamma \geq 1$ is $%
L_{11}=1-0.821\Gamma $. The inset in Fig.2 shows the $L_{11}-L_{12}$ values,
extracted from $S_{11}(k)$ and $S_{12}(k)$, corroborating the expected
behavior.

3. {\it Dielectric response matrix.} The inverse of the static dielectric
response matrix $\eta (k)=\varepsilon ^{-1}(k)$ can be expressed in terms of
the structure functions as $\eta _{ij}(k)=\delta _{ij}-\beta n \varphi
_{il}(k)S_{lj}(k)$. It determines the total potential $\Phi _i$ in the two
layers when either of them is perturbed by an external potential $\hat{\Phi}%
_i$: $\Phi _i=\eta _{ij}\hat{\Phi}_j$.

At small layer separation the effective dielectric function 
$det \varepsilon_{ij}$ has the features of the dielectric
function of an isolated 2D layer (see Fig. 4) with the
characteristic anti-screening behavior for $k<k_*$ ( $\sim 25$). This
behavior becomes qualitatively different precisely at the layer separation
where $L_{11}-L_{12}$ changes from positive to negative 
(see inset in Fig. 4)\cite{KGunp}: $%
\varepsilon _{11}$ and $\varepsilon _{12}$ develop alternatively
anti-screening and screening domains.

We can compare our results concerning the critical values of $d/a$ at the
phase transition boundaries with those of Goldoni and Peeters 
(GP)\cite{GP95}
into our $d/a$ values
Table I shows the comparison. The $d/a$ [$(ii)\rightarrow (iii)$] value has
been taken along the liquid-solid phase boundary at the lowest $\Gamma $
value ($\Gamma =118$) in Fig. 8 of their paper. GP claim a wide range of
stability ($0.79<d/a<1.10$) for the staggered square lattices. We don't see
this happening: the gradual transformation into a rhombic structure seems to
follow immediately the formation of the square lattice. The expected first
order character of the (iv) - (v) transition is masked in the present
description. Neither is there a clear indication of the transition taking
place at $\varphi =69.6^{\circ }$ predicted by GP) rather than at 
$\varphi=60^{\circ}$ the indicator would be a slight decrease in the 
$\rho _i$ values
accompanying the detectable slight increase in the $Q_i$ values). The
possibility for substitutional disorder is not part of the GP lattice model
which allows for structural changes only, although, as our analysis reveals,
it would be, for a finite temperature, a determining factor of the phase
structure of the system in the $d/a<0.5$ domain.

In summary, we have calculated, the two-body functions, structure functions 
and the
static dielectric matrix of a strongly correlated electronic bilayer liquid
through the classical HNC approach. We have found a series of dramatic
changes in the liquid structure as the layer separation varies from $d/a=0$
to $d/a=3$. These changes can be brought into correspondence with the
structural transformations and with substitutional order-disorder
transformations in the underlying lattice structure. This behavior seems to
be unparalleled in other strongly correlated liquid systems. We have also
shown how the structure functions and the static dielectric matrix
reflect these structural changes.

This work has been partially supported by NSF Grant PHY-9115714. Part of the
work was done while GK was visiting at the University of California at San
Diego and is grateful to Tom O'Neil for his hospitality. We would like to
thank Dan Dubin for the discussions and for making his unpublished data
available and to Ken Golden of the University of Vermont for a series of
useful conversations. 

\ \\CAPTIONS FOR FIGURES. 

Fig. 1. Variation of the structure of the two-body functions with increasing
layer separations $d/a$ at $\Gamma $ $=80$: (a) $g_{11}(r)$; (b) $g_{12}(r)$.

Fig. 2. Characteristic variation with layer separation of the first, second
and third maxima of $g_{11}(r)$ and $g_{12}(r)$: (a) positions $R_I$, 
$R_{II}$
(b) associated coordination
numbers $\rho _{II}$, $\rho _{II}$, $\rho _{III}$ and $\sigma _I$, $\sigma
_{II}$, $\sigma _{III}$. The insets show the five principle lattice
structures and the value $g_{12}(r)$ at $r=0$.

Fig. 3. Variation of the structure functions with increasing layer
separations $d/a$ for $\Gamma =20$ for $S_{11}(k)$ and $S_{12}(k)$. Note 
that the coordinates of $S_{11}$ and $S_{12}$ are shifted with respect 
to each other.
 
Fig. 4. Diagonal and off-diagonal elements of the static dielectric matrix
for different layer separations (a) $d/a=0.05$, ($L_{11}>L_{12}$) (b) $%
d/a=0.1$ ($L_{11}<L_{12}$) at $\Gamma =10$. The insets show $det
\varepsilon_{ij}$ and $L_{11}-L_{12}$, the difference between diagonal and
transinverse compressibilities.

CAPTION FOR TABLE

Phase boundaries ((i): simple triangular; (ii/a): substitutionally
disordered staggered quadrangular; (ii/b): substitutionally ordered
staggered quadrangular; (iii): staggered square; (iv): staggered rhombic;
(v): staggered triangular)

Table:

\begin{tabular}{|c|c|c|c|}
\hline
Phases &  & $d/a$ &  \\ \hline
& This work &  & GP \\ \hline
$(ii/a)\rightarrow (ii/b)$ & $\sim 0.5$ &  & --- \\ \hline
$(ii)\rightarrow (iii)$ & $0.7\sim 0.8$ &  & $0.79$ \\ \hline
$(iv)\rightarrow (v)$ & $1.5\sim 2.0$ &  & $1.87$ \\ \hline
\end{tabular}

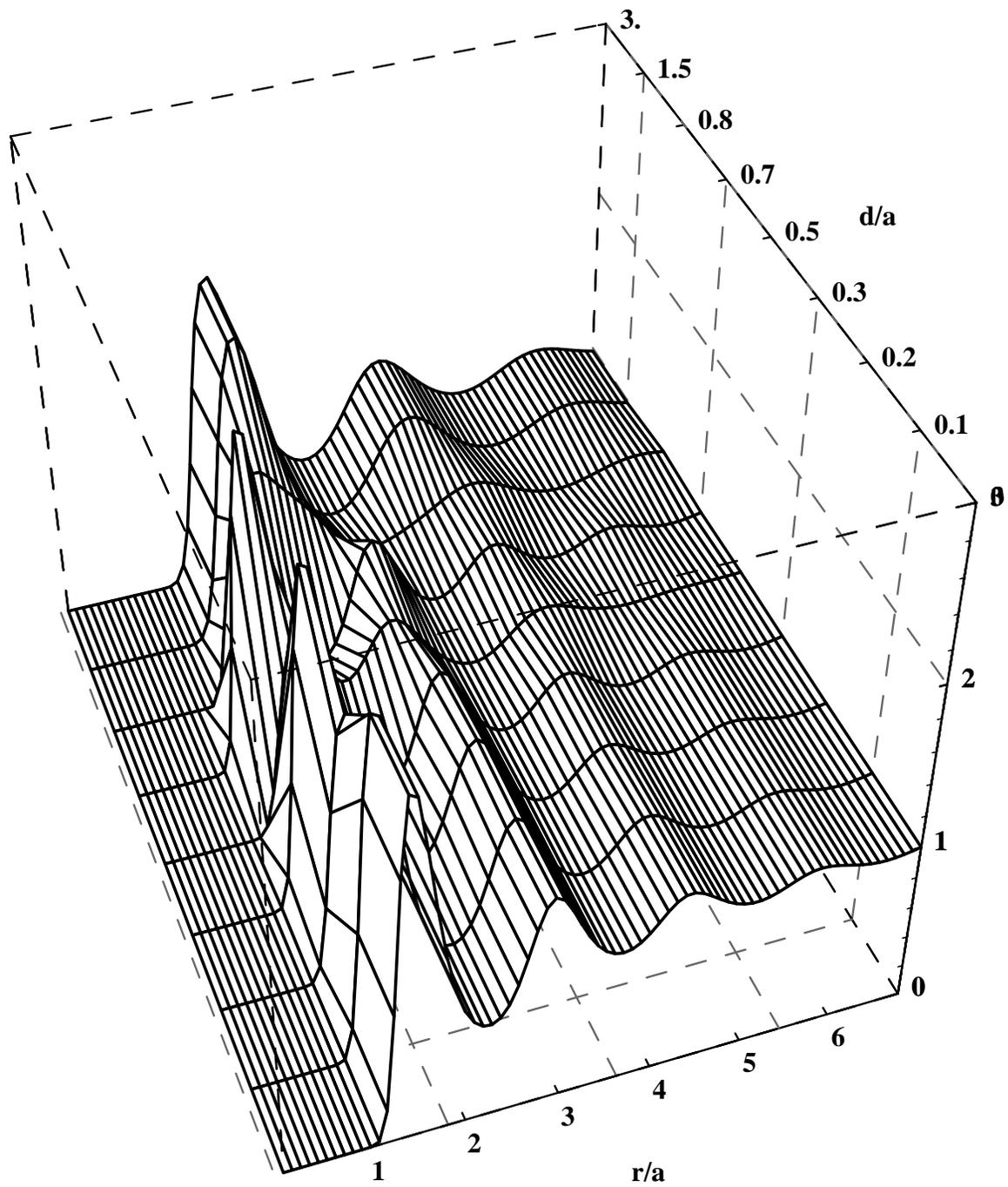
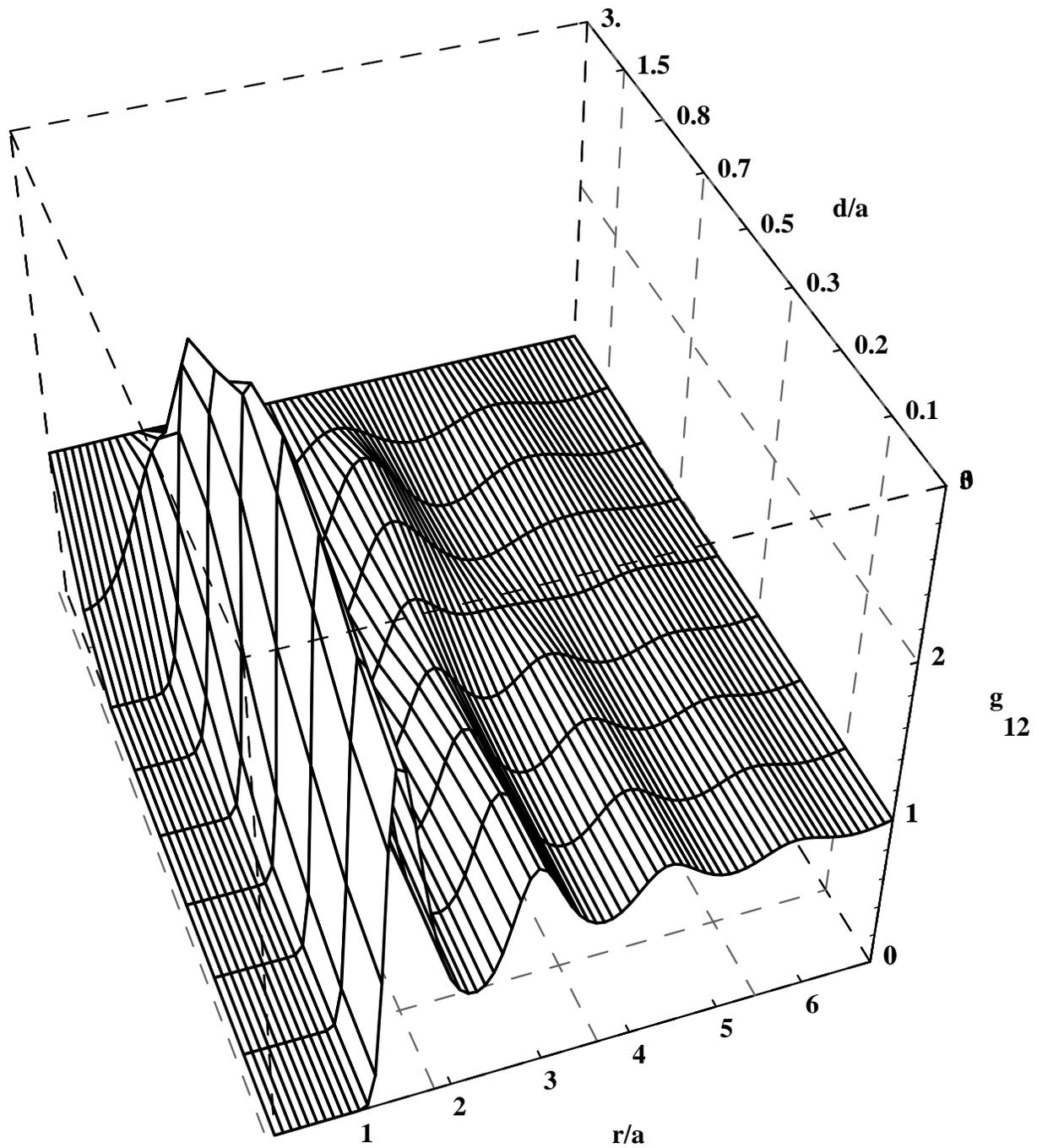
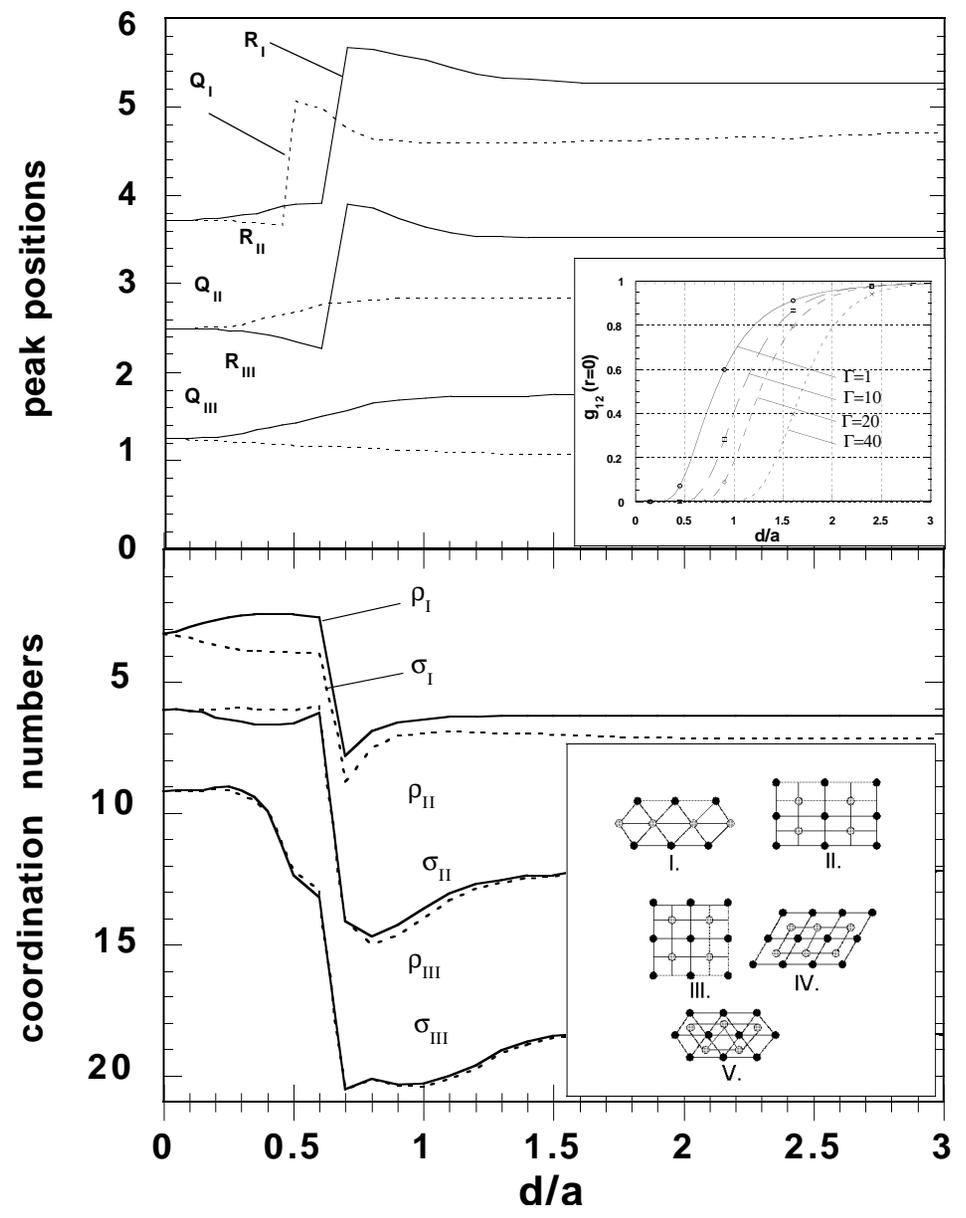
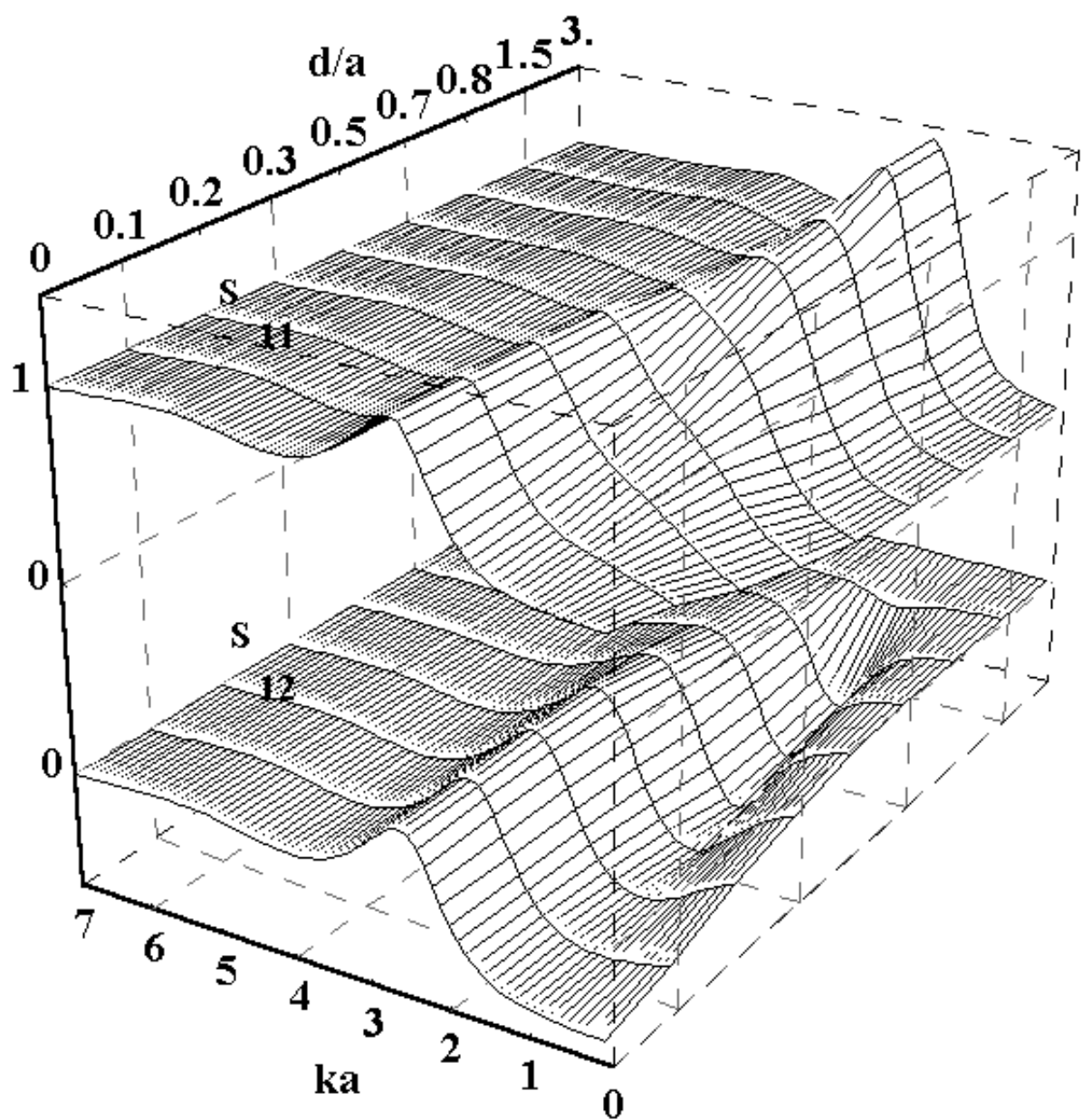
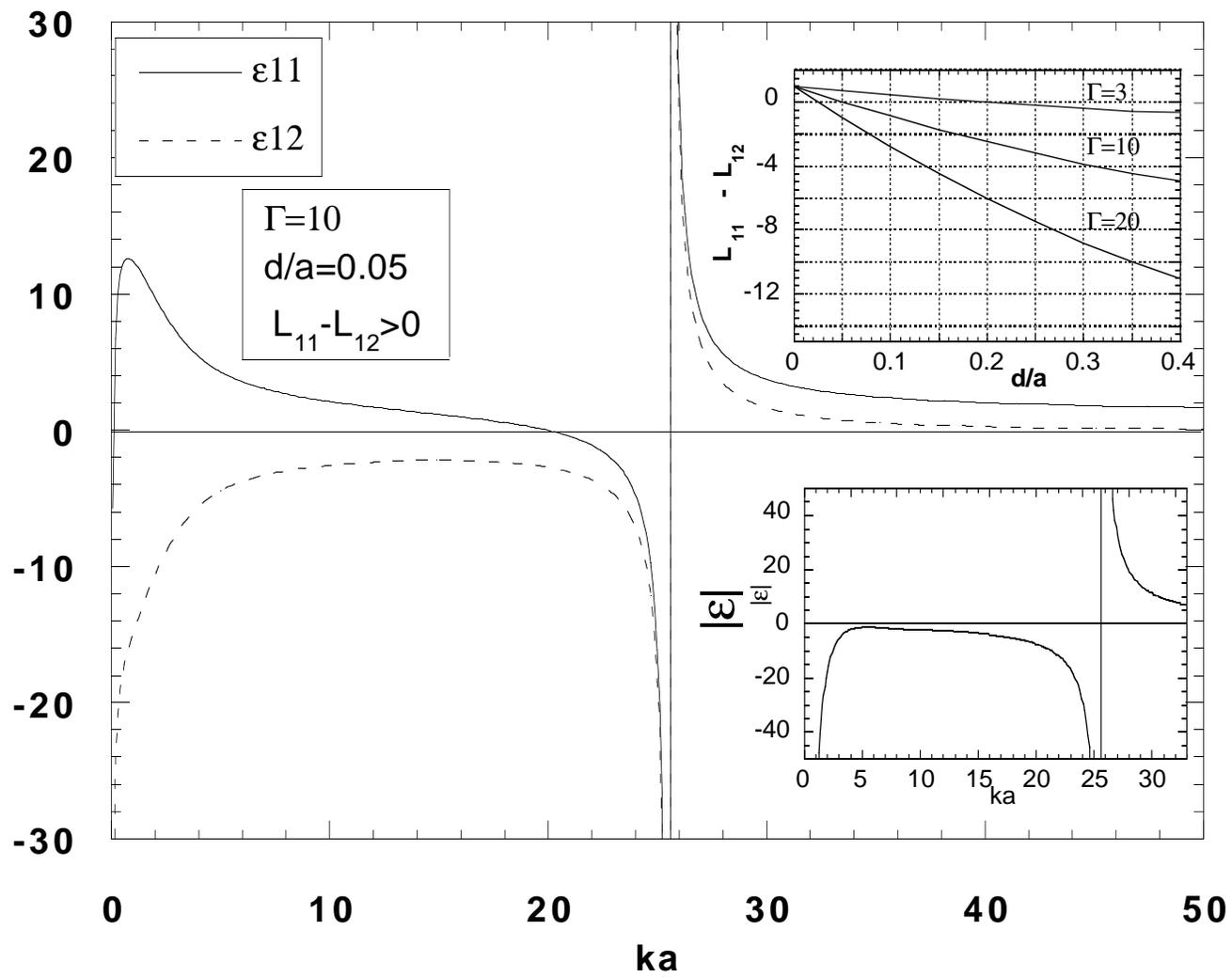
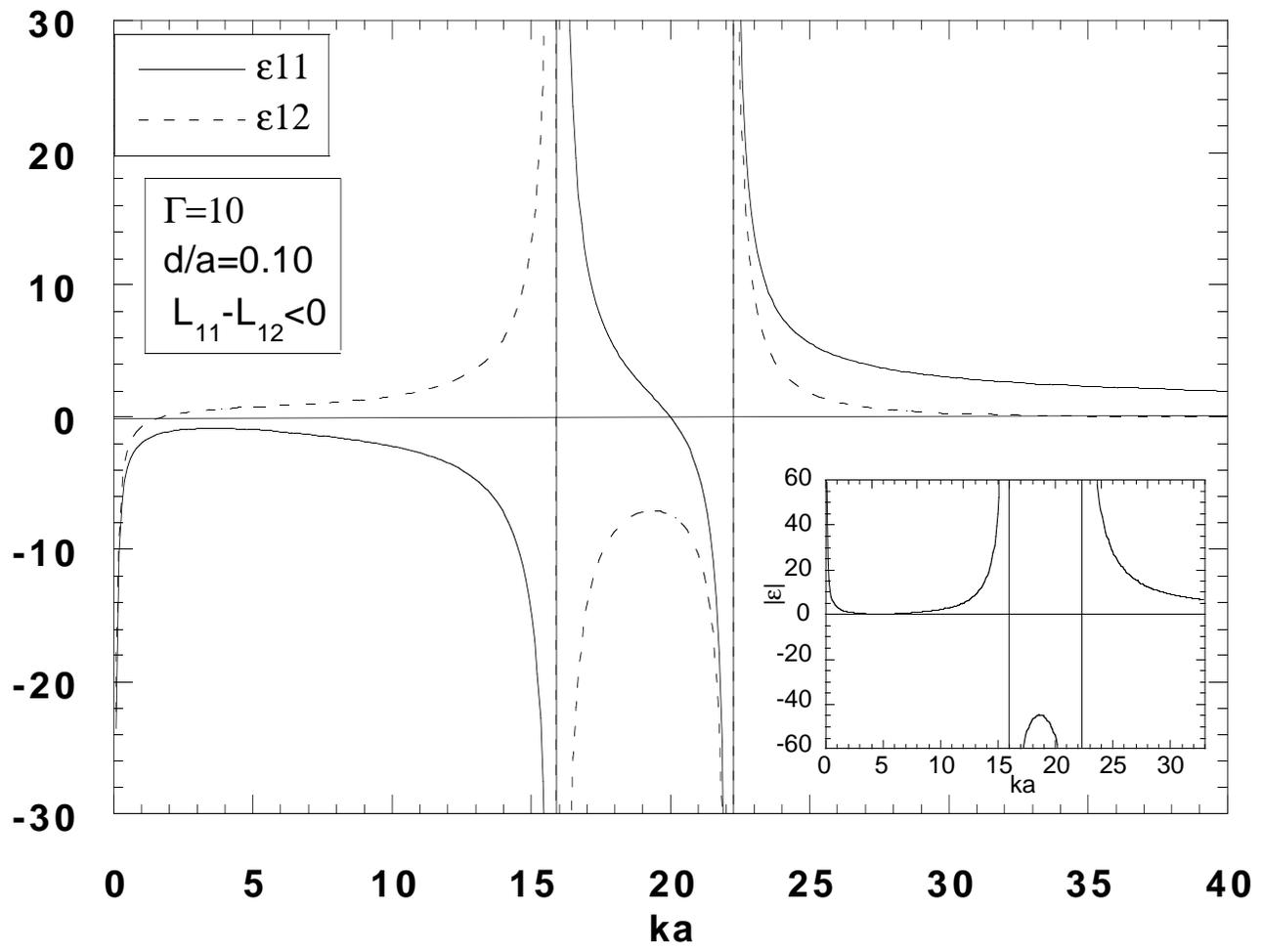
\end{document}